\documentclass[conference]{IEEEtran}
\IEEEoverridecommandlockouts
\usepackage{cite}
\usepackage{amsmath,amssymb,amsfonts}
\usepackage{algorithmic}
\usepackage{graphicx}
\usepackage{textcomp}
\usepackage{xcolor}
\usepackage{eso-pic}

\usepackage[pdftex]{hyperref}
\usepackage{tikz}

\newcommand\copyrighttext{%
    \footnotesize \textcopyright 2022 IEEE. Personal use of this material is permitted.
    Permission from IEEE must be obtained for all other uses, in any current or future
    media, including reprinting/republishing this material for advertising or promotional
    purposes, creating new collective works, for resale or redistribution to servers or
    lists, or reuse of any copyrighted component of this work in other works.
    DOI: \href{https://doi.org/10.1109/BigData55660.2022.10020688}{https://doi.org/10.1109/BigData55660.2022.10020688}}
\newcommand\copyrightnotice{%
    \begin{tikzpicture}[remember picture,overlay]
        \node[anchor=south,yshift=10pt] at (current page.south) {\fbox{\parbox{\dimexpr\textwidth-\fboxsep-\fboxrule\relax}{\copyrighttext}}};
    \end{tikzpicture}%
}

\def\BibTeX{{\rm B\kern-.05em{\sc i\kern-.025em b}\kern-.08em
    T\kern-.1667em\lower.7ex\hbox{E}\kern-.125emX}}
\begin{document}

\title{Leveraging Reinforcement Learning for \\ Task Resource Allocation in Scientific Workflows}

    \author{

        \IEEEauthorblockN{Jonathan Bader, Nicolas Zunker, Soeren Becker, Odej Kao}

        \IEEEauthorblockA{
            \{firstname.lastname\}@tu-berlin.de,
            Technische Universität Berlin, Germany
        }
  
    }

\maketitle
\copyrightnotice

\begin{abstract}
        Scientific workflows are designed as directed acyclic graphs (DAGs) and consist of multiple dependent task definitions.
They are executed over a large amount of data, often resulting in thousands of tasks with heterogeneous compute requirements and long runtimes, even on cluster infrastructures.
In order to optimize the workflow performance, enough resources, e.g., CPU and memory, need to be provisioned for the respective tasks.
Typically, workflow systems rely on user resource estimates which are known to be highly error-prone and can result in over- or underprovisioning.
While resource overprovisioning leads to high resource wastage, underprovisioning can result in long runtimes or even failed tasks.

In this paper, we propose two different reinforcement learning approaches based on gradient bandits and Q-learning, respectively, in order to minimize resource wastage by selecting suitable CPU and memory allocations.
We provide a prototypical implementation in the well-known scientific workflow management system Nextflow, evaluate our approaches with five workflows, and compare them against the default resource configurations and a state-of-the-art feedback loop baseline.  
The evaluation yields that our reinforcement learning approaches significantly reduce resource wastage compared to the default configuration.
Further, our approaches also reduce the allocated CPU hours compared to the state-of-the-art feedback loop by 6.79\% and 24.53\%.
    \end{abstract}


\section{Introduction}\label{sec:INTRO}
The paradigm of scientific workflows is used in many domains, such as genomics, physics, or remote-sensing~\cite{lehmannFORCENextflowScalable2021,rettelbach2021quantitative,bader2021tarema,baderReshi2022IPCCC}.
Scientific Workflow Management Systems (SWMS), e.g., Nextflow~\cite{nextflow}, Pegasus~\cite{pegasus}, or Saasfee~\cite{saasfee}, help scientists to compose, execute and monitor their large-scale data analysis workflows in a reproducible manner~\cite{bader2022towards}.
These SWMS submit task instances consecutively to the resource manager, e.g., Kubernetes or Slurm.
To select a fitting node for each task, the resource managers rely on resource estimates, usually provided by the user.
However, user configurations are known to be inaccurate and highly error-prone~\cite{witt2019feedback,witt2019learning}, leading to a wastage of assigned resources or even the failure of a task due to insufficient resources.
For instance, assigning too much memory, although resulting in a successful task execution, might prevent further parallel tasks and, thus, unnecessarily decrease the overall throughput.
In contrast, insufficient memory allocations can result in memory bottlenecks, i.e., memory swapping or even failed executions~\cite{will2022get,will2022ruya}.
Considering that commodity clusters often consist of heterogeneous components~\cite{cpuperformance,bader2022towards,baderStypRekowski2022ICDMW,baderLotaruLocallyEstimating2022}, this issue is even further aggravated.

Depending on the cluster infrastructure available to the scientist,  different optimization objectives can be considered~\cite{tovar2017job}:
Reducing the difference between assigned and used resources, the resource wastage, might be a desirable objective for shared or highly utilized cluster infrastructures.
Contrary, minimizing the workflow makespan might be the selected objective in cases where hard user deadlines exist and resource efficiency is subordinate.

In order to tackle this situation, we propose two reinforcement learning approaches based on gradient bandits and Q-learning to size task resources with the goal of reducing the wastage of scientific workflow tasks.
Therefore, we examine how to define actions, state spaces, policies, and reward functions for the reinforcement learning agents.
We implemented our methods in the well-known SWMS Nextflow and compared them experimentally to a user-based default configuration and a state-of-the-art approach, using five real-life scientific workflows from the popular nf-core repository~\cite{ewels2020nf}.
Our evaluation shows that both of our reinforcement learning agents significantly reduce CPU and memory wastage compared to the workflows' default resource configuration, while at the same time reducing the overall amount of allocated CPU hours compared to a state-of-the-art feedback loop baseline.

\emph{Outline}. The remainder of the paper is structured as follows.
Section~\ref{sec:RELATED_WORK} discusses related work.
Section~\ref{sec:APPROACH} explains our two reinforcement learning approaches in detail.
Section~\ref{sec:EVALUATION} evaluates the reinforcement learning agents and compares them to the workflows' default configurations and a state-of-the-art feedback loop.
Section~\ref{sec:CONCLUSION} summarizes and concludes our paper.

\section{Related Work}\label{sec:RELATED_WORK}
First, we cover literature related to the general problem of task sizing in scientific workflows.
Second, we discuss approaches that use reinforcement learning to allocate resources.

\subsection{Sizing of Scientific Workflow Tasks}

Witt et al.~\cite{witt2019feedback} propose a feedback-based resource allocation system for scheduling scientific workflows.
Their approach uses an online feedback loop to improve resource usage predictions and thus more accurately sizes resources to the tasks of a scientific workflow. 
The authors examine two approaches to estimate peak memory usage.
The percentile predictors, e.g., P50, uses the memory usage of a task's percentile memory consumption as its prediction.
The second method is based on a linear regression that estimates peak memory consumption of the workflow tasks based on the input data size.
The predictions are then offset by the standard deviation between the predicted and the true peak memory usage.

The memory allocation problem is also addressed in~\cite{witt2019learning}.
Here, the authors use the input data size to predict the peak memory usage and train a prediction model that minimizes resource wastage instead of the prediction error.
They train a linear model that takes the data size as an input to maximize the memory allocation quality.
Further, the authors examine different failure handling strategies.

In contrast to the presented work, we use reinforcement learning for resource allocation and also consider the sizing of CPU cores.

Torvar et al.~\cite{tovar2017job} present a job sizing strategy for tasks in scientific workflows which either minimizes the wastage or maximizes the throughput. 
For the initial task sizing attempt, the authors use the probabilities of peak resource values from the historical traces.
Their goal is to minimize the sum of probabilities of resource peaks where the resource peak is greater than the allocated resource value.
If the so-defined first resource allocation leads to an under-prediction and the task fails, Tovar et al. first assign the maximum ever observer memory. 
If this allocation also fails, they assign the entire resources of the most powerful node to the task.

The scope of our paper is similar since we also address the sizing of CPU and memory for tasks.
However, our reinforcement learning approaches include the exploration of the environment to find possibly better allocations than provided in the historical traces, while their approach applies a minimization based on historical observations.
Further, once a task fails, we propose a more selective sizing strategy to avoid high resource wastage.

\subsection{Resource Allocation with Reinforcement Learning}

SmartYARN~\cite{smartYarn} is an extension of the well-known resource manager YARN and uses reinforcement learning to optimize efficiency and performance simultaneously.
More specifically, the authors apply a q-learning approach to application samples to learn a policy.
One aspect of their paper is the trade-off between reducing costs by using fewer resources and the need to decrease runtime by using more resources.
They used the performance of the application under a certain resource configuration as the state space of the agent. 
The action space includes increasing or decreasing one unit of CPU or memory or keeping the previous allocation. 
Similar to our approach, the authors consider the trade-off between cost and performance when assigning resources. 
However, one difference is that SmartYARN optimizes individual and independent jobs, whereas scientific workflows consist of task dependencies expressed in a directed acyclic graph.

Mao et al. propose DeepRM~\cite{deepRM}, a resource manager using deep reinforcement learning. 
DeepRM uses a policy gradient reinforcement learning algorithm combined with a neural network.
The state space consists of the cluster's resources and the resource requirements of arriving jobs encoded as "images" to feed into the neural network. 
The actions available to the agent are choosing a job and allocating resources to it. 
Through the training process, DeepRM learned a policy to keep small amounts of resources available to quickly assign any small jobs and maximize the throughput of small jobs.
While DeepRM also considers which resource to schedule a job, our focus is solely on task sizing.

VCONF~\cite{vconf} addresses a similar problem in the scope of resource allocations of virtual machines by using reinforcement learning. 
Their approach models the problem as a continuing discounted Markov Decision Process. 
The states are triples of CPU time credits, virtual CPUs, and memory. At the same time, the available actions are either increasing, decreasing, or leaving the allocations, with only one resource allowed to be changed per action.
By modeling states, the agent is able to simulate or predict the reward it can expect from previously unseen action-state pairs, whereas the classic q-learning approach, used in our paper, requires the agent to experiment with each action-state pair. 
Therefore, the model-based agent can learn much faster and enter its exploitation phase earlier than a static agent, which must laboriously explore the action-state space, especially when there are large state spaces. 
The key difference to our approach is the scope of scheduling virtual machines while we consider the amount of resources to allocate.
Further, the authors use model-based reinforcement learning to approximate a value function and reduce the exploration time.

\section{Approach}\label{sec:APPROACH}
In this section we first present an abstract overview of our approach in the context of a scientific workflow environment.
Second, our gradient bandit approach, which is composed of a CPU bandit and a memory bandit, is presented.
Finally, we explain our Q-Learning approach that combines both, CPU and memory sizing.

\subsection{Approach Overview}
\label{sub:overview}

\begin{figure*}
\centering
    \includegraphics[width=0.65\textwidth]{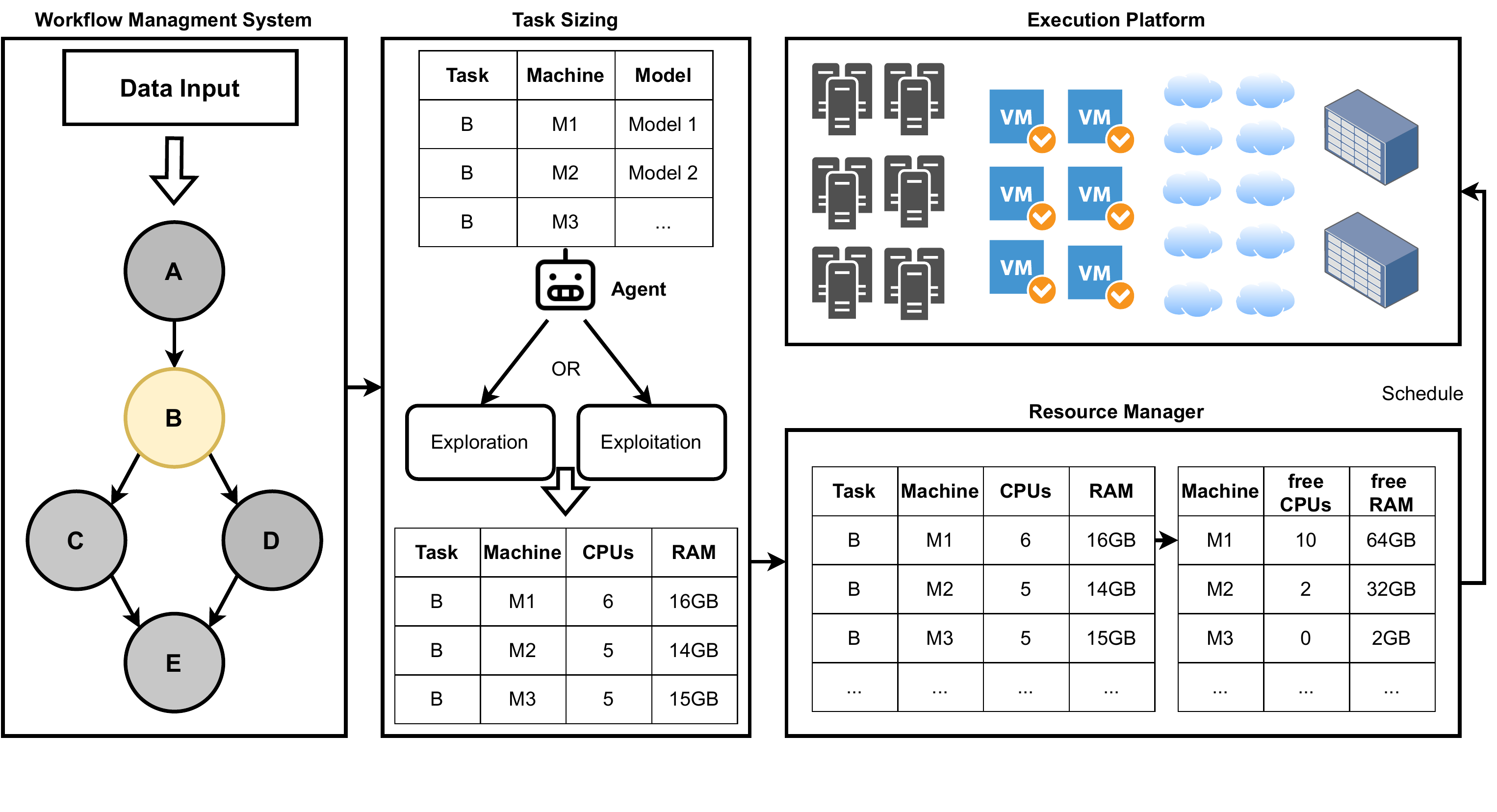}
    \caption{Overview of our approach in the context of a typical scientific workflow environment.} \label{fig:arch}
\end{figure*}

Figure~\ref{fig:arch} provides an overview of our reinforcement learning task sizing approach.
%The workflow management system submits ready task instances one by one.
Instead of directly submitting ready tasks to the resource manager, in our approach, the SWMS is extended by reinforcement learning agents that define the CPU and memory allocations for the respective tasks.
Since real-world scenarios are often heterogeneous, i.e., task and infrastructures characteristics, we employ a reinforcement learning agent instance for each abstract task-machine combination.
Therefore, one agent for each of these combinations exists, whereby each agent decides individually whether to further explore resource combinations or to exploit the currently best one.
Thus, the resource manager receives a list of task-machine pairs and their proposed resource allocations which in turn can be used to create an online scheduling plan.

\subsection{Gradient Bandits}
\label{sub:gradient_bandit}

Gradient Bandits learn a preference $H_t(a)$ for each action $a \in A$, where A is the space of actions and $t$ a discrete time step.
The relative preference $H_t(a)$ influences the probability $\pi_t(a)$ to chose the specific action $a$, where $\pi_t(a)$ is defined as follows:

\begin{equation}
\pi_{t}(a) = \frac{e^{H_t(a)}}{\sum_{q \in A} e^{H_t(q)}}
\end{equation}

and $H_{t=0}(a) = 0 \ \forall a$.

Then, the preference value for an action $a$ at the discrete time step $t+1$ is defined as follows~\cite{sutton2018reinforcement}:

\begin{equation} 
	H_{t+1}(a) = \begin{cases}
		H_t(a) + \alpha (R_t - \bar R)(1 - \pi_{t} (a)) &\text{if $a = {a_t}$}\\
		H_t(a) - \alpha (R_t - \bar R) \pi_{t} (a) &\text{else}
	\end{cases}
\end{equation}

where $\alpha$ is the step-size and $R_t$ the reward from the respective reward function.
The choice of $a$ can be translated to the selection of a specific CPU or memory configuration.
In the following, we will discuss the selection of $\alpha, A$ and the reward function for the CPU and memory bandit.

\paragraph{CPU Bandit}

The CPU Bandit has a restricted set of actions to choose from which are based on the number of CPU cores available to the system. The bandit then learns how many CPU cores to allocate to its task. The reward function is defined as follows:

\begin{equation}
\label{bandit1_reward}
reward = -t\cdot (1+cpus - \frac{cpu_{usage}}{100})
\end{equation}

where the $cpu_{usage}$ is a value in percent of the number of CPU cores used by the task, e.g a usage of two and a half CPU cores would result in a $cpu_{usage}$ of 250\%. 
Further, $cpus$ is the number of CPU cores assigned to the task and $t$ is a tasks' runtime.
We punish the agent with a negative reward equal to the amount of time it ran plus the unused CPU time (execution time multiplied by the number of unused CPU cores).
Our main idea behind this is to provide an incentive for the agent to minimize the CPU idle time.
Including the negative task runtime, we encourage the agent to reduce the amount of time taken for the task to complete.
For instance, when a task's actual usage matches the assigned usage, i.e., $cpus - \frac{cpu_{usage}}{100}= 0$, we use the tasks' execution time as its penalty.
Without a concept of time, the agent would always allocate the least amount of resources possible, leading to very slow task executions.

Since the range of our reward function~\ref{bandit1_reward} is  ${[-t \cdot (cpus+1), -t]}$, our expected reward is unbounded since the execution time can be arbitrary large.
Without having a mean reward and a standard deviation, the step size $\alpha$ can not be chosen as a constant scalar value anymore.
Instead, we use a function which incorporates the average time that it takes the task to complete.
Through this, tasks with longer average runtimes are given smaller step sizes. 
Therefore, we selected the value for $\alpha$ as $\alpha = \frac{1}{avg(time)}$

\paragraph{Memory Bandit}
\label{sub:mem_bandit}

Our memory bandit assigns an amount of memory, where different possible memory allocations define the available actions to the agent.
Since we want to assign the memory in bytes, we have to limit the number of available actions to a manageable amount.
Thus, we split the amount of memory $m$ from the default configuration into $n$ chunks each of size $c= \frac{m}{n}$.
Then, the bandit can choose between $a$ actions, where each action $a$ assigns $a \cdot c$ amounts of memory to the task. 
Since users often overestimate the memory resources needed to avoid a task failure~\cite{witt2019learning,witt2019feedback}, most of the time $a \in [1,n]$ holds and an initial configuration for $a$ with $a < n$ encourages memory efficiency.

The biggest difference to the previously presented CPU bandit is that a task with insufficient memory assigned will either be stopped by the resource manager or result in possible slow memory swapping ~\cite{witt2019feedback,witt2019learning, will2022get}. 
Therefore, our reward function for the memory bandit includes a penalty for such cases that is defined as follows:

\begin{equation}
\label{bandit2_reward}
reward = \begin{cases}
     -2\cdot mem_{asg}/c& \text{task ran out of memory}\\
    -1 \cdot chunks_{unused}              & \text{otherwise}
\end{cases}
\end{equation}

where $mem_{asg}$ defines the memory assigned, $chunks_{unused} = \frac{mem_{asg} - peak_{rss}}{c} $ and $peak_{rss}$ is the peak memory usage in bytes during the execution.
The presented reward function is bounded in the range of $[- 2n, 0]$.
Therefore, a fixed step size of $\alpha = 1/n$ can be chosen for the reinforcement learning agent.

Whenever a task fails due to too little memory, a new allocation is tried that must be greater than the failed one.
If the bandit picks a smaller allocation, it is ignored.
Then, either the new allocation is doubled or the default configuration for that task is used in case double the new allocation is less than the failed allocation.

\subsection{Q-learning Agent}
\label{sec:q_agent}
In addition to the gradient approach, we also provide a Q-learning based method which combines the selection of CPU and memory allocations.
Here, we again utilize one Q-learning agent per abstract task. 
We set the Q-learning agent’s state to the current allocation of CPU and memory for a given task.
As set of actions, the approach either increases or decreases the CPU and memory allocation, or does nothing. 
Again, we have to limit the state space, i.e., the minimum and maximum number of CPU cores and memory as well as the amount by which the reinforcement learing agents can increment or decrement the resources.
We set the states for the Q-learning agent accordingly to the previously defined state spaces of the CPU and the memory bandits.

Our Q-learning agent incorporates both aspects, CPU and memory sizing.
Therefore, only a single reward function exists, defined as follows:

\begin{equation}
\label{q_agent_1_reward}
reward = -cpu_{waste} \cdot \frac{t}{avg\_t} \cdot mem_{waste}
\end{equation}
where 
\begin{gather}
\label{foo}
cpu_{waste} = max(0.1,cpu_{unused})\\
mem_{waste} = mem_{asg} \cdot (1 - min(0.75,mem_{use})
\end{gather}

Again, $cpu_{unused} = cpus - \frac{cpu_{usage}}{100}$, $mem_{asg}$ refers to a task's assigned memory, and $mem_{use}$ is the value of a task's peak memory divided by the memory assigned to it.
Additionally, $t$ is defined as a task's runtime and $avg\_t$ is determined at runtime based on the historical average execution time for the specific task.

This function is effectively a product of the number of unused CPU cores, the slow-down factor, and the amount of unused memory. 
The $max$ and $min$ functions are used to set an artificial floor for the penalty incurred by the unused CPU cores and unused memory.
Tasks utilizing more than 90\% of the available CPU cores are given the same reward as tasks which use exactly 90\%. 
Additionally the floor for unused memory is capped at 25\% in a similar manner. 
Thus, we discourage the agent from allocating insufficient memory to prevent out-of-memory failures.

The average task execution time is already incorporated into the reward function.
Therefore, there is no need to consider it for the step sizes as in the gradient bandit (Section~\ref{sub:gradient_bandit}).
Consequently, the issues associated with the step size are avoided.

\section{Evaluation}\label{sec:EVALUATION}

In this section, we present our experimental evaluation of the proposed approaches and our prototypical implementation.
\subsection{Experimental Setup}
\label{sec:testing}
%\paragraph{Prototype Implementation} 
We implemented our reinforcement learning approaches as an extension for the SWMS Nextflow~\cite{nextflow}.
By default, Nextflow gathers runtime metrics from the Linux process stats (ps).
To enable an online sizing, we extended the workflow system through a monitoring extension that stores task-resource traces at runtime in a database.
This information is then used by our sizing agents, together with their states, actions, and reward function, to find an optimal resource configuration for the task before submitting it to the resource manager. 

\paragraph{Cluster Setup and Evaluation Workflows}

We evaluate our approach on a machine with an AMD EPYC 7282 CPU (16C32T, 2.8GHz base frequency), 128GB of DDR4 memory, and two 1TB SSDs in a RAID 1 configuration.

We selected five real-world workflows from the popular nf-core repository~\cite{ewels2020nf}.
All workflows were used with five different test data input profiles, leading to 25 different combinations in total.
The five workflows used are the following: \raisebox{.5pt}{\textcircled{\raisebox{-.9pt} {1}}} \textit{eager} - a pipeline for genomic NGS sequencing data~\cite{yates2021reproducible}, \raisebox{.5pt}{\textcircled{\raisebox{-.9pt} {2}}} \textit{mhcquant} - a workflow for quantitative processing of data dependent (DDA) peptidomics data, \raisebox{.5pt}{\textcircled{\raisebox{-.9pt} {3}}} \textit{nanoseq} - which is an analysis pipeline for Nanopore DNA/RNA sequencing data, \raisebox{.5pt}{\textcircled{\raisebox{-.9pt} {4}}} \textit{viralrecon} - which can be  used to perform assembly and intra-host/low-frequency variant calling for viral sample, and \raisebox{.5pt}{\textcircled{\raisebox{-.9pt} {5}}}  \textit{metaboigniter} - a pipeline for pre-processing of mass spectrometry-based metabolomics data.

First, we run each workflow ten times without our reinforcement learning agents, using the default resource allocation specified in the workflow declaration.
The collected traces serve as the input for the agents as well as the second baseline.
We tested the bandits 50 times for each workflow and the Q-learning agent 100 times, in order to provide for a more extensive state-action space for the latter.
%The decision to test the Q-learning more times than the bandits is due to a more extensive state-action space, resulting in a longer training phase. 

Additionally to the default configuration baseline, we include a feedback loop approach based on the work from \cite{tovar2017job} Tovar et al. and \cite{witt2019feedback} Witt et al., which was presented in Section~\ref{sec:RELATED_WORK}.
During ten training runs, the feedback loop approach first assigns the maximum number of CPU cores and memory available for each task.
In the next runs, the tasks are assigned their average resource usage recorded during the training and the mean plus the standard deviation of their usage.
In case of a  failure, the task is retried with the maximum amount of memory ever used by that task during the training time, and should it fail again then it is retried with double that value.

For the final comparison, the last ten runs of each of the agents are compared with ten runs which used the default configurations and the last ten runs which applied the feedback loop approach.

\subsection{Results}

In our first experiment we compare the CPU allocation of our reinforcement learning approaches with the baselines and depict the resulting resource wastage.
The second experiment presents the differences regarding the memory allocation performance.

\paragraph{CPU Allocation}

\begin{figure}[h]
    \centering
        \includegraphics[width=\columnwidth]{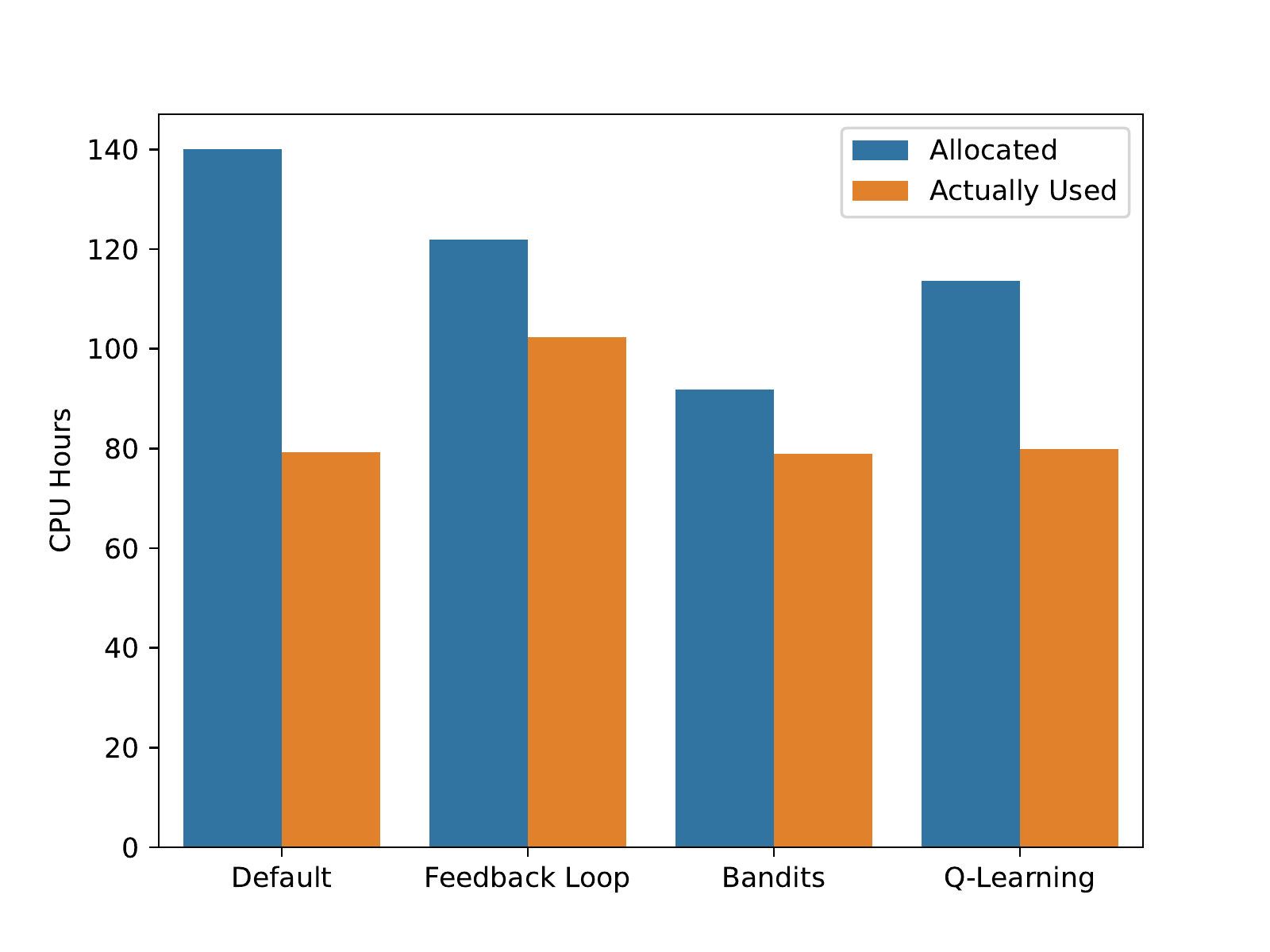}
        \caption{Allocated and actually used CPU hours over all workflows for the baselines and the reinforcement learning approaches.}
        \label{fig:cpu_waste}
\end{figure}

Figure~\ref{fig:cpu_waste} compares the allocated and the actually used number of CPU hours accumulated over all five workflows.
The difference between allocated and actually used resources is the wastage.
One can see that the default configuration yields the highest wastage with around 61 CPU hours.
Our bandit approach achieves the lowest wastage with around 10 CPU hours.
The Q-learning agent leads to 34 CPU hours wastage and the feedback loop to 20 CPU hours wastage.
Noticeable is that the default configuration, our bandits, and our Q-Learning approach all use around 80 CPU hours.
However, in contrast, the feedback loop uses 102 CPU hours.
We tested this configuration several times to validate this outlier.
This behaviour indicates that an increase in CPU resources does not necessarily result in a linear decrease of runtime, but rather converges at a specific point. Still, the feedback loop approach often selects configurations with a high amount of CPU cores which explains the greater actual used CPU hours in the plot.

\begin{figure}[h]
    \centering
        \includegraphics[width=\columnwidth]{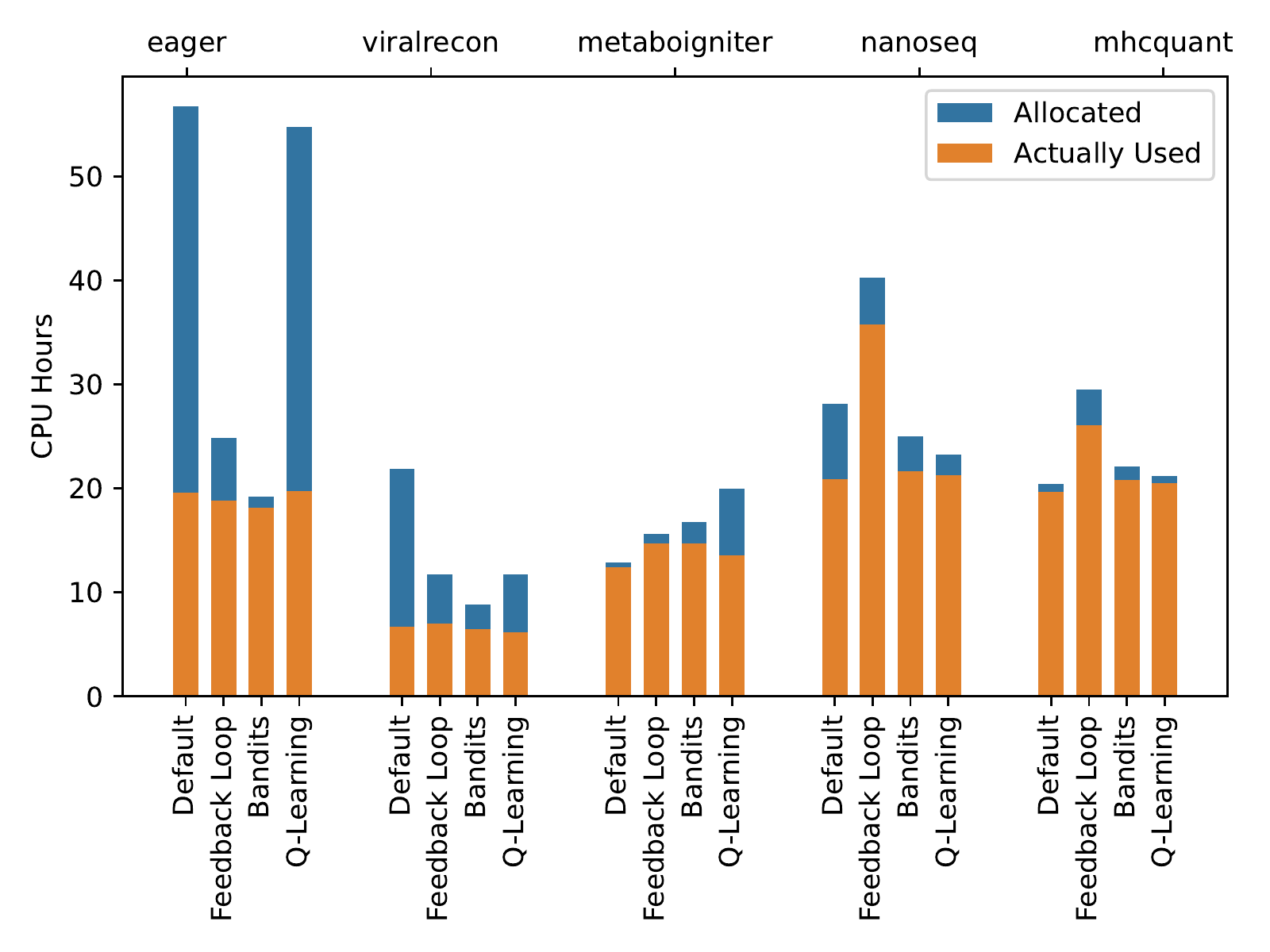}
        \caption{Allocated and actually used CPU hours per workflow for the baselines and the reinforcement learning approaches.}
        \label{fig:cpu_waste_wfs}
\end{figure}

Figure~\ref{fig:cpu_waste_wfs} shows the allocated and actually used number of CPU hours on a per-workflow basis.
We can observe that for the metaboigniter and the mhcquant workflow, the default configuration provides settings that lead to low out-of-the-box wastage, i.e., the user-estimates are quite accurate.
In contrast, the default configuration is rather bad for the viralrecon and, especially, for the eager workflow.
Further, the bandit approach yields good results for all workflow profiles.
Additionally, Figure~\ref{fig:cpu_waste_wfs} depicts the overall results of the Q-learning agent, showing that the wastage is mostly due to the eager workflow, while the Q-learning agent performs well for all others.  
In three out of five workflows, the feedback loop performs worst after the default configuration.

\paragraph{Memory Sizing}

\begin{figure}[h]
    \centering
        \includegraphics[width=\columnwidth]{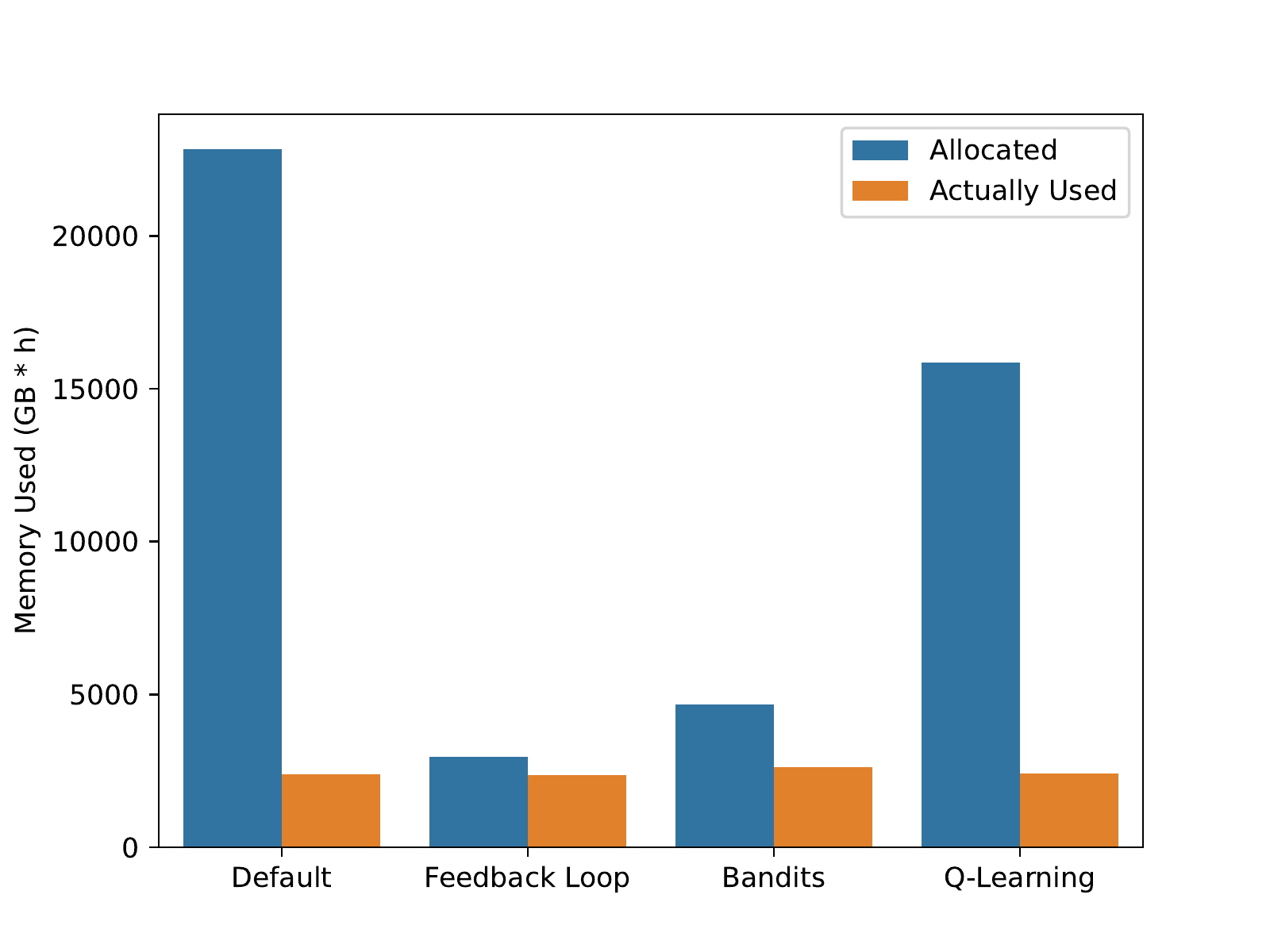}
        \caption{Allocated and actually used memory over all workflows for the baselines and the reinforcement learning approaches.}
        \label{fig:mem_waste}
\end{figure}

Figure~\ref{fig:mem_waste} shows the allocated and actually used memory for the two baselines and our two reinforcement learning approaches over all tested workflows in log scale.
The default configuration yields a high wastage of around 20456 GBh of memory.
In contrast, the lowest wastage, 601 GBh, is achieved by the feedback loop baseline, leading to an average usage of 90\%.
The bandits approach is the second-best performing one with a wastage of 2060 GBh.
Our Q-learning agent shows a wastage of 13418 GBh.
This is significantly worse than the feedback loop and the bandits approach but still outperforms the default configuration.

\begin{figure}[h]
    \centering
        \includegraphics[width=\columnwidth]{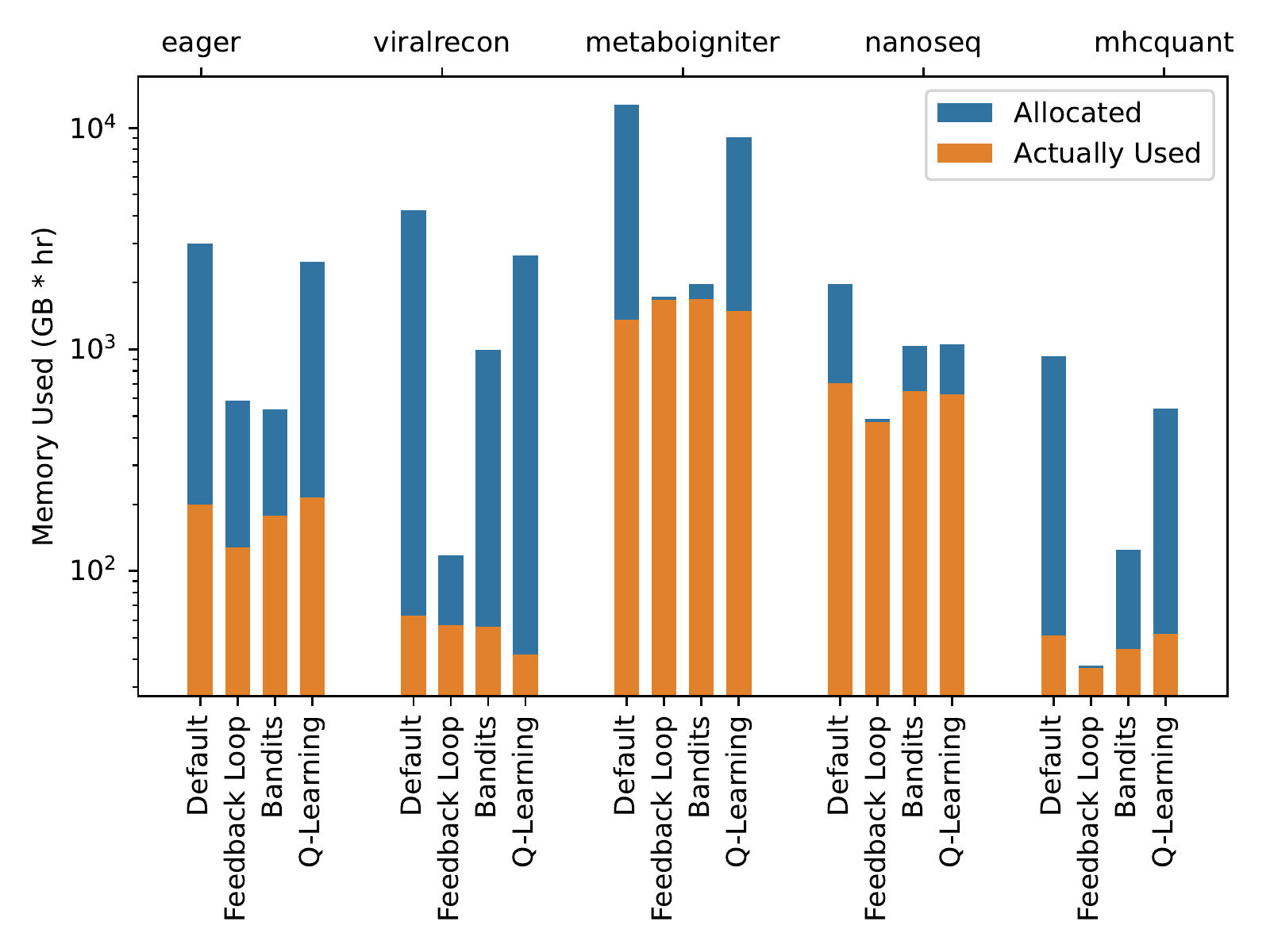}
        \caption{Allocated and actually used memory per workflow for the baselines and the reinforcement learning approaches.}
        \label{fig:mem_waste_wfs}
\end{figure}

Finally, Figure~\ref{fig:mem_waste_wfs} describes the actually and totally assigned GBh of memory for each workflow individually.
Four out of five times, the feedback loop achieves the lowest wastage.
The bandits achieve the lowest wastage for one workflow and four times the second-lowest wastage.
In contrast to the default CPU wastage, where a relatively low CPU wastage is achieved for some workflows, the default configuration for memory always leads to the highest wastage.

\section{Conclusion}\label{sec:CONCLUSION}
In this paper, we proposed two reinforcement learning approaches using gradient bandits and Q-learning to allocate task resources in scientific workflows. 
Our approach examines the challenges in defining actions, state spaces, policies, and reward functions for resource allocation and provides solutions for this problem.

We implemented a prototype of our reinforcement learning strategies into the workflow management system Nextflow and conducted an evaluation with five different workflows from the popular nf-core repository.
The results show that we are able to significantly reduce CPU and memory wastage compared to the default configuration.
Further, although leading to greater memory wastage compared a state of the art feedback loop approach, the proposed methods are able to outperform it in terms of allocated CPU resources.

Currently, our agents have to treat tasks independently and do not consider their interdependencies.  
In future work, we aim to integrate our reinforcement learning agent into the scheduler of a workflow management system to enrich the agents with a broader context.
Therewith, the agents could, for example, assign slightly fewer resources than usual to a task in order to use the slack time for increased efficiency.

\subsection*{Acknowledgments}
{\small Funded by the Deutsche Forschungsgemeinschaft (DFG, German Research Foundation) as FONDA (Project 414984028, SFB 1404).}

    \bibliographystyle{IEEEtran}
    \bibliography{./references}

\end{document}